\documentclass[10pt]{article}

\usepackage{amsmath}   
\usepackage{amsfonts}  
\usepackage{graphicx}  

\input amssym.def
\input amssym
\newcommand{\be}{\begin{equation}}
\newcommand{\ee}{\end{equation}}

\begin{document}

\vspace*{2.5cm}
\renewcommand{\thefootnote}{\fnsymbol{footnote}}
\noindent{\Large\bf Rotation and Pseudo-Rotation}

\vspace*{0.8cm} \noindent{\bf Nikolai V.
Mitskievich\footnote{Physics Department, C.U.C.E.I., University of
Guadalajara, Guadalajara, Mexico, e-mail nmitskie@gmail.com}~ and
H\'ector Vargas-Rodr{\'\i}guez\footnote{Physics Department,
C.U.C.E.I., University of Guadalajara, Guadalajara, Mexico, e-mail
hv{\underline{\phantom{a}}}8@yahoo.com}}

\vspace*{0.5cm} \hspace*{1.5cm}\begin{minipage}[t]{9.5cm}

\hrulefill

\noindent Eigenvectors of stress-energy tensor (the source in
Einstein's equations) form privileged bases in description of the
corresponding space-times. When one or more of these vector fields
are rotating (the property well determined in differential
geometry), one says that the space-time executes this rotation.
Though the rotation in its proper sense is understood as that of a
timelike congruence (vector field), the rotation of a spacelike
congruence is not a less objective property if it corresponds to a
canonical proper basis built of the just mentioned eigenvectors.
In this last case, we propose to speak on pseudo-rotation. Both
properties of metric, its material sources, and space-time
symmetries are considered in this paper.

\hrulefill


\end{minipage}

\begin{flushright}
\{  See the (mixed) English--Spanish--Russian poem\\
   dedicated to Alberto A. Garc\'ia D\'iaz in [15]\\
     as it was published in Gen. Relat. and Gravit.\}
\end{flushright}

\vspace*{0.5cm}

\noindent{\large\bf 1. INTRODUCTION}

\vspace*{0.5cm}

\noindent One seems to know quite well what is the rotation
incorporated into the metric tensor. But there is an alternative
way, sometimes used in describing, for example, the Kerr metric in
a frame ``without rotation'', but with an alternative combination
of $d\varphi$ and $dt$ (now in this succession), see, {\em e.g.},
the textbook by Misner, Thorne, and Wheeler [7], Exercises 33.3
and 33.4. We call this choice of frame as that with
`pseudo-rotation', being studied below, alongside with rotation,
using some typical examples of non-empty space-times (the Kerr
space-time is not so much appropriate since it corresponds to a
vacuum, thus in the absence of material rep\`eres, with Killing
vectors only remaining as a possible tool). This non-emptiness
means that the space-time is filled by some sort(s) of matter
(electromagnetic field, fluids, {\em etc.}), in other words, there
actually is a stress-energy tensor as a source in Einstein's
equations. The concrete structure of this tensor not only
determines the nature of matter under consideration, but also
({\em via} its eigenvectors) influences upon the properties of
space-time, for example, through a rotation.

In this paper (published in [15]) we consider rotation and its
counterpart, pseudo-rotation. The first one corresponds to
rotation of a timelike congruence, and the second (whose
importance is widely underestimated), to rotation of a spacelike
congruence. When these congruences (equivalently, vector fields)
reflect objective properties of a physical system ({\em e.g.},
they are Killing vectors of the space-time, or eigenvectors of the
stress-energy tensor), they obviously have equal logical footing
and importance. Any approach based on Killing vectors is however
quite restrictive: if a space-time possessing isometries is
superimposed with even small exterior perturbation (say, a
gravitational wave comes from a faraway source), its symmetries
disappear, although this cannot mean that the rotation so abruptly
ceases to exist. This vulnerability of exact mathematical
symmetries calls for extreme care when one intends to draw from
them a working physical definition.

In the next Section we give prominence to rotation of perfect
fluids using their field-theoretic description. In this case, the
rotation gains on pseudo-rotation, since the former is so easily
visualizable as a rotation about a spatial axis (the axis of a
pseudo-rotation may be timelike, not only spacelike). Moreover,
the four-velocity of a perfect fluid is equivalent not only to the
timelike basis vector of the proper tetrad, but also to the
intensity of the 2-form field, thus, {\em via} its field equation,
it should be closely related to the specific mechanism which
introduces rotation in the theory (we emphasize that rotation
plays in the theory of the 2-form field, at least formally, a
similar r\^ole to that of the sources in the electromagnetic and
gravitational theories). Nevertheless, there are nice and
important exact solutions in general relativity (also
Einstein--Maxwell fields) with pseudo-rotation and a combination
of rotation and pseudo-rotation, and one cannot ignore them,
especially because of their fundamental simplicity and symmetric
structure. In Sections 3 and 5 we consider two such examples,
while in Section 4 we find that pseudo-rotation automatically
appears in the well-known Kerr--Newman space-time between the
event and Cauchy horizons.

Below we are working in four space-time dimensions with signature
\linebreak $(+,-,-,-)$, Greek indices being four-dimensional. The
Ricci tensor is $R_{\mu\nu}={R^\alpha}_{\mu\nu\alpha}$, thus
Einstein's equations take the form
$R_{\mu\nu}-\frac{1}{2}Rg_{\mu\nu}= -\varkappa T_{\mu\nu}$.

\newpage

\noindent{\large\bf 2. ROTATING  FLUIDS  IN  FIELD-THEOR\-ETIC
DESCRIPTION}

\vspace*{0.5cm}

\noindent Perfect fluids can be conveniently described with use of
the Lagrangian formalism, especially in the absence of rotation
[9, 10]. In this case they are represented via the 2-form field
potential $B=\frac{1}{2!}B_{\mu\nu}dx^\mu\wedge dx^\nu$, the
respective field intensity being $G=dB=\frac{1}{2}
B_{\mu\nu;\lambda} dx^\lambda\wedge dx^\mu\wedge dx^\nu$
($B_{[\mu\nu;\lambda]} \equiv B_{[\mu\nu,\lambda]}$, and
$G_{\lambda\mu\nu}=B_{\lambda\mu,\nu}+B_{\mu\nu,\lambda}+
B_{\nu\lambda,\mu}$) whose invariant $J=\ast(G\wedge\ast G)$ (we
shall also denote $\ast G= \tilde{G}$) is used in constructing the
fluid Lagrangian density ${\frak L}=\sqrt{-g}L(J)$. Here the Hodge
star $\ast$ denotes, as usual, a generalization of the dual
conjugation applied to Cartan exterior forms: with an $r$-form
$\alpha=\alpha_{\nu_1\dots\nu_r} dx^{\nu_1}\wedge\cdots\wedge
dx^{\nu_r}$, it yields a $(4-r)$-form $\ast\alpha$ with the
components $(\ast\alpha)_{\nu_1\dots \nu_{4-r}}=\frac{1}{r!}
E_{\nu_1\dots\nu_{4-r}\nu_{5-r} \dots\nu_4}\alpha^{\nu_{5-r}
\dots\nu_4}$ where $E_{\varkappa \lambda\mu\nu}= \sqrt{-g}
\epsilon_{\varkappa \lambda\mu\nu}$ and $E^{\varkappa\lambda\mu
\nu}=-(1/\sqrt{-g}) \epsilon_{\varkappa \lambda\mu\nu}$ are co-
and contravariant axial skew rank-4 tensors, $\epsilon_{\varkappa
\lambda\mu\nu}= \epsilon_{[ \varkappa\lambda\mu\nu]}$,
$\epsilon_{0123}=+1$ being the Levi-Civit\`a symbol ({\it cf.}
somewhat other notations in [8]).

The reason why this description of perfect fluids is valid, is
simply the fact that the stress-energy tensor of a 2-form field is
\be T^\beta_\alpha=2J\frac{dL}{dJ}b^\beta_\alpha
-L\delta^\beta_\alpha \label{1} \ee where \be
b^\beta_\alpha=\delta^\beta_\alpha- u_\alpha u^\beta,
b^\beta_\alpha u^\alpha=0= b^\beta_\alpha u_\beta, ~
u=J^{-1/2}\tilde{G}.\label{2} \ee When $u$ is timelike ($u\cdot
u=+1$, as we above supposed it to be), we come to the usual
perfect fluid whose (arbitrary) equation of state is determined by
the dependence of $L$ on its only argument, $J$ (see [9], [10])
[however when it is spacelike, the `fluid' is tachyonic (see for
some details [12], subsection 3.2)]. Since $b^\beta_\alpha$ is the
projector on the (local) subspace orthogonal to the congruence of
$u$, the latter is an eigenvector of the stress-energy tensor with
the eigenvalue $(-L)$, $T^\beta_\alpha u^\alpha=-Lu^\beta$, while
any vector orthogonal to $u$ is also eigenvector, now with the
three-fold eigenvalue $2J\frac{dL}{dJ} -L$. This is the property
of the stress-energy tensor of a perfect fluid possessing the
proper mass density $\mu$ and pressure $p$ (in its local rest
frame): \be\label{mu_p} \mu=-L, ~ ~ p=L-2J\frac{dL}{dJ}. \ee

Below we consider perfect fluids characterized by the simplest
equation of state \be \label{eqstate} p=(2k-1)\mu \ee (the
frequently used notation is $2k=\gamma$) which correspond to the
Lagrangian $L=-\sigma|J|^k$, $\sigma>0$. In a four-dimensional
spacetime, the important special cases are: the incoherent dust
($p=0$) for $k=1/2$, intrinsically relativistic incoherent
radiation ($p=\mu/3$) for $k=2/3$, and hyperrelativistic stiff
matter ($p=\mu$) for $k=1$.

However the 2-form field equation which follows from the above
Lagrangian, \be\label{3} \left(\sqrt{-g}\frac{dL}{dJ}
G^{\lambda\mu\nu}\right)_{,\nu}=0 ~ \Longleftrightarrow ~
d\left(J^{1/2}\frac{dL}{dJ}u\right)=0, \ee only means that the
$\tilde{G}$ (equivalently, $u$) congruence is non-rotating. To
describe a rotating fluid, one has to introduce in (\ref{3}) a
non-zero right-hand side. This, in a sharp contrast to the usual
equations of mathematical physics ({\it cf.}, for example,
electrodynamics), {\em cannot then be interpreted as a usual
source term} (this was stressed in [12]): its meaning essentially
is to indicate the presence of rotation ($u\wedge du\neq 0$). To
this end it is necessary to consider one more field which we call
the Machian one, a 3-form field $C$ with the intensity $W=dC$ (see
[9, 10]). In terms of $L(K)$, $K=-(1/4!)W_{\kappa\lambda\mu\nu}
W^{\kappa\lambda\mu\nu}=\tilde{W}^2$, its equations reduce to
\be\label{4} \left(\sqrt{-g}\frac{dL}{dK}
W^{\kappa\lambda\mu\nu}\right)_{,\nu}=0 ~ \Rightarrow ~
K^{1/2}\frac{dL}{dK}=\mbox{const.} \ee We use also the duality
relations $B\!\stackrel{\mu\nu}{*}=\frac{1}{2}
E^{\mu\nu\alpha\beta} B_{\alpha\beta}$, $G_{\lambda\mu\nu}=
\tilde{G}^\kappa E_{\kappa\lambda\mu\nu}$, $W_{\kappa\lambda
\mu\nu}=\tilde{W}E_{\kappa\lambda\mu\nu}$. Moreover,
${B\!\stackrel{\mu\nu}{*}}_{;\nu}\equiv -(\ast G)^\mu$.

Since we were confronted with the no rotation property of perfect
fluid when the rank 2 field was considered to be free, the only
remedy now is to introduce a non-trivial ``source'' term in the
$r=2$ field equations, thus to consider the non-free field case
or, at least, to include in the Lagrangian a dependence on the
rank 2 field potential $B$. The simplest way to do this is to
introduce in the Lagrangian density dependence on a new invariant
$J_1= -B_{[\kappa\lambda}B_{\mu\nu]}B^{[\kappa\lambda}B^{\mu\nu]}$
which does not spoil the structure of stress-energy tensor,
simultaneously yielding a ``source'' term (thus permitting to
destroy the no rotation property) without changing the divergence
term in the $r=2$ field equations. We shall use below three
invariants: the obvious ones, $J$ and $K$, and the just introduced
invariant of the $r=2$ field {\em potential}, $J_1$. Then \be
\label{6} B_{[\kappa\lambda}B_{\mu\nu]}=-\frac{2}{4!}
B_{\alpha\beta} B\!\stackrel{\alpha\beta}{\ast}E_{\kappa\lambda
\mu\nu}. \ee Thus $J_1^{1/2}= 6^{-1/2} B_{\alpha\beta}
B\!\stackrel{\alpha\beta} {\ast}$. In fact, $J_1=0$, if $B$ is a
simple bivector ($B=a\wedge b$, $a$ and $b$ being 1-forms). This
{\em cannot however annul} the expression which this invariant
contributes to the $r=2$ field equations: up to a factor, it is
equal to $\partial J_1^{1/2}/\partial B_{\mu\nu}\neq 0$. Thus let
the Lagrangian density be \be\label{7} {\frak L}=
\sqrt{-g}\left(L(J)+M(K) J_1^{1/2}\right), \ee so that the $r=2$
field equations take the form ({\em cf.} (\ref{3})) \be\label{x}
d\left(\frac{dL}{dJ} \tilde{G}\right)= \sqrt{\frac{2}{3}}M(K)B ~
\Leftrightarrow ~
\left(\sqrt{-g}\frac{dL}{dJ}{G^{\alpha\beta\nu}}\right)_{,\nu}=
\sqrt{-g}
\sqrt{\frac{2}{3}}M(K)B\!\stackrel{\alpha\beta}{\ast}.\ee In their
turn, the $r=3$ field equations ({\em cf.} (\ref{4})) yield the
first integral \be\label{y} J_1^{1/2} K^{1/2}\frac{dM}{dK}=
\mbox{const}\equiv 0 \ee (since $J_1=0$). We know from [9, 10]
that $K$ (hence, $M$) {\em arbitrarily} depends on the space-time
coordinates, if only the $r=3$ field equations are taken into
account, and the Machian field $K$ has to be essentially
non-constant.

The stress-energy tensor which corresponds to the new Lagrangian
density (\ref{7}), automatically coincides with its previous form
(\ref{1}), since $J_1=0$. For a perfect fluid with the equation of
state $p=(2k-1)\mu$, one finds $L=-\sigma J^k$, thus
$T^\beta_\alpha=-2kLu_\alpha u^\beta+(2k-1)L \delta^\beta_\alpha$.
Then the traditional perfect fluid language is obviously related
with that of the $r=2$ and $r=3$ fields: \be\label{z} \left.
\begin{array}{l} \displaystyle {\mu=-L=\sigma J^k, ~ ~
\tilde{G}^\mu =\Xi\delta^\mu_t, ~ ~ \Xi=\frac{1}{\sqrt{g_{00}}}
\left(\frac{\mu}{\sigma}\right)^{1/(2k)}},\\
\displaystyle{G=dB=d\left(\frac{\sqrt{3/2}}{M(K)}\right)\wedge
d\left( \frac{dL}{dJ}\tilde{G}\right)} \end{array} \right\} \ee
({\em cf.} (\ref{x})). The function $M$ depends arbitrarily on
coordinates; thus one can choose its adequate form using the last
relation without coming into contradiction with the dynamical
Einstein--Euler equations.

When one describes a fluid in its proper basis, $u=J^{-1/2}\tilde
G=\theta^{(0)}$, the rotation of the fluid's co-moving reference
frame is defined as $\omega=\ast\left(\theta^{(0)}\wedge d
\theta^{(0)}\right)=J^{-1}\ast(\tilde G\wedge d\tilde G)$. Let us
assume $\theta^{(0)}=e^\alpha(dt+fd\phi)$ where $\alpha$ and $f$
are functions of coordinates (usually determined {\em via}
Einstein's equations), {\em cf.} the examples of metrics
considered in the next Sections, though in these examples are
treated Einstein--Maxwell fields and still not the perfect fluid
solutions. It is inevitable to conclude that the field theoretic
approach to perfect fluids automatically gives hints and even
concrete relations (often having a simple algebraic form) imposed
upon these and other functions characterizing the metric tensor
and the 2-form field, as well as the Machian one. This makes it
possible to substantially simplify the treatment of Einstein's
equations. The purpose of this paper is not to come into further
details of such calculations, and we shall return to them in other
publications (for some simple examples see [12]).

\newpage

\noindent{\large\bf 3. A  SIMPLE  ELECTROVACUUM  SPACETIME\\
WITH ROTATION  AND  PSEUDO\-ROTATION}

\vspace*{0.5cm}

 \noindent We now consider a special case
($\Phi(u)=C/\sqrt{ \varkappa}=$ const.) of the conformally flat
null Einstein--Maxwell field (32.103) in [4], whose metric
obviously takes the Kerr--Schild form $ds^2=dt^2- d\rho^2-\rho^2
d\tilde\varphi ^2 -dz ^2+ \frac{C^2}{4}\rho^2(dt-dz)^2$, as well
as the cylindrically symmetric forms with both rotation and
pseudo-rotation \be\label{4a} ds^2=\left(dt +\frac{C}{2}  
\rho^2d\varphi\right)^2-d\rho^2-\rho^2d\varphi^2-\left(dz+
\frac{C}{2}\rho^2 d\varphi\right)^2\ee and \be\label{5a} 
ds^2=(d\tilde t+C x d y)^2-d x^2-d y^2-(d\tilde z+C x d y)^2.\ee
The corresponding natural orthonormal tetrads are: for (\ref{4a}),
\be\label{6a} \theta^{(0)}=d t+\frac{C}{2}\rho^2 \, d\varphi,  
~~~~ \theta^{(1)}=d \rho,~~~~ \theta^{(2)}=\rho d \varphi,~~~~
\theta^{(3)}=dz+\frac{C}{2}\rho^2 \, d\varphi,\ee and for
(\ref{5a}), \be\label{7a} \tilde \theta^{(0)} 
=d\tilde t+Cx\, dy,~~~~ \tilde \theta^{(1)}=d x,~~~~ \tilde
\theta^{(2)}=d y,~~~~ \tilde \theta^{(3)}=d\tilde z+Cx \, d y.\ee
The relations between coordinates (those with a tilde and without
it, as well as $\rho$, $\phi$ and $x$, $y$) are obvious.

It is remarkable that this space-time admits seven independent
Killing vectors given here in the coordinates of (\ref{4a}), but
in the basis (\ref{6a}): \be \label{Kv1} \xi_{[1]}=\theta^{(0)},
~~ \xi_{[2]}=-\theta^{(3)}, ~~ \xi_{[3]}=
\frac{C}{2}\rho^2\left(\theta^{(0)}-\theta^{(3)}\right)-\rho
\theta^{(2)},\ee \be \label{Kv2} \xi_{[4]}= C\rho\cos\varphi
\left(\theta^{(0)}-\theta^{(3)}\right)-
\sin\varphi\theta^{(1)})-\cos\varphi\theta^{(2)},\ee \be
\label{Kv3} \xi_{[5]}=C\rho \sin\varphi\left(\theta^{(0)}-
\theta^{(3)}\right)+\cos\varphi\theta^{(1)}-\sin\varphi
\theta^{(2)},\ee \be \label{Kv4} \xi_{[6]}= \cos[C(t-z)-\varphi]
\theta^{(1)}+ \sin[C(t-z)-\varphi]\theta^{(2)},\ee \be \label{Kv5}
\xi_{[7]}=\sin[C(t-z)-\varphi] \theta^{(1)}-\cos[C(t- z)
-\varphi]\theta^{(2)}.\ee Contravariant Killing vectors in the
coordinate frame satisfy the following non-trivial commutation
relations:
$$ \begin{array}{l}
{[\xi_{[1]},\xi_{[6]}]=-C\xi_{[7]}},\\
{[\xi_{[1]},\xi_{[7]}]=C\xi_{[6]}},\\
{[\xi_{[2]},\xi_{[6]}]=C\xi_{[7]}},\\
{[\xi_{[2]},\xi_{[7]}]=-C\xi_{[6]}},\\
{[\xi_{[3]},\xi_{[4]}]=-C\xi_{[5]}},
\end{array}
~~~~~~~~
\begin{array}{l}
{[\xi_{[3]},\xi_{[5]}]=-C\xi_{[4]}},\\
{[\xi_{[3]},\xi_{[6]}]=C\xi_{[7]}},\\
{[\xi_{[3]},\xi_{[7]}]=-C\xi_{[6]}},\\
{[\xi_{[4]},\xi_{[5]}]=C(\xi_{[1]}+\xi_{[2]})},\\
{[\xi_{[6]},\xi_{[7]}]=C(\xi_{[1]}+\xi_{[2]})},
\end{array}
$$
while $\xi_{[1]}\cdot\xi_{[1]}=1$, $\xi_{[3]}\cdot\xi_{[3]}=-
\rho^2$, and the five other spacelike Killing vectors are unitary
($\xi\cdot\xi=-1$). It is worth mentioning that $\xi_{[1]}$ and
$\xi_{[2]}$ are rotating and pseudo-rotating respectively, the
first around the axis $\theta^{(3)}$ and another, `around'
$\theta^{(0)}$, with one and the same magnitude of `angular
velocity',
$$
\omega=\frac{1}{2}*(\xi_{[1]}\wedge d
\xi_{[1]})=\frac{C}{2}\tilde\theta^{(3)}, ~~~
\varpi=\frac{1}{2}*(\xi_{[2]}\wedge d
\xi_{[2]})=\frac{C}{2}\tilde\theta^{(0)}.
$$
Another remarkable property of the space-time under consideration
is that its metric can be expressed exclusively in terms of the
Killing covectors: \be \label{dsKil} ds^2=\xi_{[1]}\xi_{[1]}-
\xi_{[2]}\xi_{[2]} -\xi_{[6]}\xi_{[6]}-\xi_{[7]}\xi_{[7]}. \ee

It might seem that the explicit form of the Maxwell field as the
source of the gravitational field of this simple electrovacuum
space-time is already known being a special case of the more
general solution given in [4], [13], but this is not exactly the
case. We show here that there is a multitude of electromagnetic
fields (in the sense of the field tensor and, of course, not only
of the potential) which yield one and the same stress-energy
tensor in the fixed four-geometry under consideration, and this is
a perfectly special case in general relativity completely beyond
the framework of the well known invariance of the stress-energy
tensor with respect to the dual conjugation of the field tensor
$F_{\mu\nu}$. Moreover, some of the seven Killing vectors being at
our disposal, when multiplied by a suitable constant coefficient,
not only satisfy the vacuum Maxwell equations in this space-time
(which is only natural due to the well known Wald theorem, see [6]
and --- for applications to the case of test electromagnetic
fields --- [2]), but give together with the geometry (the
gravitational field) of the space-time, self-consistent solutions
of the Einstein--Maxwell equations, and this is not only one
solution, but {\em a multitude} of self-consistent solutions in
one and the same space-time. One of us (N.~M.), in collaboration
with J. Horsk\'y, developed and applied a new method of purposeful
constructing exact self-consistent Einstein--Maxwell fields using
Killing vectors of seed space-times [3]. Naturally, this method
led always to generalizations of these seed space-times, the
Killing vector having generated exact perturbations of seed
geometries. Now we find that in this new special case, the Killing
vector (in fact, four of them simultaneously), up to a constant
factor directly related to the parameter in the metric tensor,
already represents the electromagnetic four-potential of this
self-consistent solution. And different Killing vectors (of these
four) form different self-consistent solutions whose
four-geometry, however, is exactly one and the same. Quite
naturally, we came to this conclusion without any intention to
find such a clear example or even look for it at all. Of course,
since the geometries created by these different fields, exactly
coincide, and Maxwell's equations are linear with respect to the
electromagnetic field, a superposition of these fields becomes
automatic, without any apparent interaction between such
electromagnetic fields. Only the constant parameter in the metric
has to be built by additive contributions of the coefficients by
individual Killing vectors.

Let an electromagnetic four-potential be proportional to a Killing
covector, $A=k\xi$. The corresponding electromagnetic field tensor
then is $F_{\mu\nu}=k(\xi_{\nu;\mu}-\xi_{\mu;\nu})=
2k\xi_{\nu;\mu}$. The identity $\xi_{\alpha;\gamma;\beta}-
\xi_{\alpha;\beta;\gamma}=\xi^\delta R_{\alpha\delta\beta\gamma}$
and the Killing equation yield \be \label{Kid} \xi_{\alpha;\beta;
\gamma}= \xi^\delta R_{\alpha\beta\gamma\delta}.\ee Due to the
structure of $F_{\mu\nu}$ and (\ref{Kid}), Maxwell's equations
take the form ${F^{\mu\nu}}_{;\nu}=-2k \xi^\nu R^\mu_\nu$, while
$F\stackrel{\mu\nu}{*}_{;\nu}=0$ follows from the Ricci
identities. Now it is clear that we have to consider only such
Killing vectors which are orthogonal to the Ricci tensor (the
electrovacuum case) which for (\ref{6a}) has components
$R^{(\mu)(\nu)} =\frac{C^2}{2}\left(\delta^\mu_0+\delta^\mu_3
\right)\left(\delta_0^\nu+\delta_3^\nu\right)$ (hence it is clear
that the scalar curvature $R=0$). Since $\left(\delta^\mu_0+
\delta^\mu_3 \right)$ (here the tetrad basis is used) is
orthogonal to the five Killing vectors $\xi_{[3]}$, $\xi_{[4]}$,
$\xi_{[5]}$, $\xi_{[6]}$, and $\xi_{[7]}$ (the combination
$\xi_{[1]}-\xi_{[2]}$ is excluded since this is an exact form thus
not producing any electromagnetic field), we have five candidates
for four-potentials (1-forms). This final proof is based on the
desired form \be \label{EEMT} T= \frac{C^2}{2\varkappa}
(\theta^{(0)} -\theta^{(3)}) \otimes(\theta^{(0)}-\theta^{(3)})
\ee of the electromagnetic stress-energy tensor (here it is worth
being mentioned that (\ref{EEMT}) has the standard canonical
structure for a null electromagnetic field ({\em cf.} [5], [14]).
In fact, only the Killing vector $\xi_{[3]}$ is not successful in
yielding the form (\ref{EEMT}), thus merely describing a test
electromagnetic field in this space-time; all other four Killing
vectors do indeed pertain to self-consistent Einstein--Maxwell
solutions involving the space-time under consideration. Some pairs
of them describe dually conjugated electromagnetic situations, and
their linear combinations (with appropriate constant coefficients)
correspond to `dual rotations', but there are also completely
different ones two of which we shall consider below.

In order to determine the electric and magnetic field vectors we
introduce a reference frame described by the monad (see [8]) which
we choose to be \be \tau= \theta^{(0)}=\xi_{[1]} \ee in order to
correspond to the rotating-pseudo-rotating basis (\ref{4a}). This
reference frame is rotating, $\omega=\frac{1}{2}*(\tau\wedge
d\tau)=\frac{C}{2} \theta^{(3)}$, but has neither acceleration,
$G=-*(\tau\wedge*d\tau)=0$, nor expansion and shear since the
rate-of-strain tensor vanishes, $D_{\mu\nu}=\frac{1}{2}
\pounds_\tau b_{\mu\nu}=0$ ({\it cf.} [8]). Here
$b_{\mu\nu}=g_{\mu\nu}- \tau_\mu\tau_\nu$ is the three-metric in
the local subspace orthogonal to the monad and $\pounds$ denotes
the Lie derivative. With respect to this reference frame we split
the electromagnetic field tensor in the electric and magnetic
(co)vectors \be E=* (\tau\wedge *F), ~~~~ B=*(\tau\wedge F).
\label{(3.2.7)} \ee

{\em The fourth Killing vector case.} The electromagnetic
four-potential, field tensor, and electric and magnetic
(co)vectors are $$ A_{[4]}=\sqrt{\frac{2\pi}{\varkappa}}C
x(dt-dz), ~~~~~ F_{[4]}=\sqrt{\frac{2\pi}{\varkappa}}
C\theta^{(1)}\wedge\left(\theta^{(0)}- \theta^{(3)}\right), $$
$$ E_{[4]}=\sqrt{\frac{2\pi}{\varkappa}}C\theta^{(1)}, ~~~~~
B_{[4]}=\sqrt{\frac{2\pi}{\varkappa}}C\theta^{(2)}. $$ Here we
have constant mutually orthogonal electric and magnetic fields
with equal magnitudes (a static pure null field). Formally, one
may say that this solution contains an electromagnetic wave whose
frequency is equal to zero.

{\em The sixth Killing vector case.} The electromagnetic
four-potential, field tensor, and electric and magnetic
(co)vectors are $$
A_{[6]}=\sqrt{\frac{2\pi}{\varkappa}}\left\{\cos[C(t-z)]dx
+\sin[C(t-z)]dy\right\}, $$
$$ F_{[6]}=\sqrt{\frac{2\pi}{\varkappa}}C\left\{
\sin[C(t-z)]\theta^{(1)}-\cos[C(t-z)]\theta^{(2)}\right\}
\wedge\left(\theta^{(0)}-\theta^{(3)}\right), $$
$$ E_{[6]}=\sqrt{\frac{2\pi}{\varkappa}}C\left\{\sin[C(t-z)]
\theta^{(1)}- \cos[C(t-z)]\theta^{(2)} \right\}, $$
$$ B_{[6]}=-\sqrt{\frac{2\pi}{\varkappa}}C\left\{
\cos[C(t-z)]\theta^{(1)}- \sin[C(t-z)]\theta^{(2)} \right\}. $$
When $C>0$, this pure null electromagnetic field represents a left
circularly polarized (positive helicity) plane monochromatic wave
with frequency $C$.

In all cases, the electromagnetic linear momentum density
(coinciding with the Poynting covector in our units) is equal to
\be S=\frac{1}{4\pi}*(E \wedge \tau\wedge B)=-
\frac{C^2}{2\varkappa} \theta^{(3)} \ee (see [8]); it is constant,
directed along the positive $z$ axis and does not depend on the
sign of $C$. The plane electromagnetic wave has its spin angular
momentum in an opposite direction to that of the angular velocity
of the reference frame. If $C$ changes its sign, then the plane
electromagnetic wave acquires negative helicity, and the relative
situation continues to be as before.

All these solutions possess a semi-cylindrical symmetry (\`a la
Wils), since the Lie derivatives with respect to the Killing
vectors of the space-time, $\pounds_{\xi}$, in general do not
annul the electromagnetic field tensor: this property holds in all
cases with respect to $\xi_{[3]}$; moreover, for $F_{[6]}$ and
$F_{[7]}$ this also occurs with respect to $\xi_{[1]}$ and
$\xi_{[2]}$.

\vspace*{1.cm}

\noindent{\large\bf 4. THE KERR--NEWMAN SPACE-TIME}

\vspace*{0.5cm}

We consider in this Section the well-known rotating space-time
filled with electromagnetic field which has well determined proper
directions (eigenvectors) rigidly connected with the field
distribution. Thus we can trace interrelations between the
material properties (the electromagnetic field visualized {\em
via} its stress-energy tensor) and their four-geometric
description (the behavior of a properly chosen tetrad).

{\em The rotating frame.} The usual rotating tetrad in the
Boyer--Lindquist coordinates is \be \label{KNrot} \theta^{(0)}=
e^\alpha(dt+afd\phi), ~ \theta^{(1)}=e^\beta dr, ~ \theta^{(2)}
=e^\gamma d\vartheta, ~ \theta^{(3)}=e^\delta \sin\vartheta d\phi
\ee where
$$
e^{2\alpha}=\frac{\Delta-a^2\sin^2\vartheta}{\rho^2},  ~ ~
e^{2\beta}=\frac{\rho^2}{\Delta}, ~ ~ e^{2\gamma}=\rho^2,
$$
$$
e^{2\delta}=\Delta\frac{\rho^2}{\Delta- a^2\sin^2\vartheta}, ~ ~
f=a\frac{r^2+a^2-\Delta}{\Delta-a^2\sin^2\vartheta}\sin^2\vartheta,
$$
$$
\Delta(r)=r^2-2Mr+a^2+Q^2, ~ ~ \rho(r,\vartheta)=r^2+a^2
\cos^2\vartheta.
$$
Hence, $e^{2(\beta-\gamma)}=\Delta^{-1}$, $e^{2(\alpha+\delta)}
=\Delta$, so that $\sqrt{-g}=\rho^2\sin\vartheta$. The
$\theta^{(0)}$ congruence obviously rotates: this can be seen as
non-vanishing of $\theta^{(0)}\wedge d\theta^{(0)}$.

One has to keep in mind that physically the rotation property of a
frame of reference is related to a {\em timelike} congruence whose
unit tangent vector is the monad $\tau$ (see [8]) denoted here by
$\theta^{(0)}$, but in the Kerr--Newman case, as this can be seen
from the above expressions, its square changes sign, at least with
$\phi$ remaining constant, when $\Delta=a^2\sin^2\vartheta$ (the
well-known static limit), thus on this surface the tetrad is
inadmissible. The situation is then similar to that of
inadmissibility of the Boyer--Lindquist coordinates on the
horizons (for simplicity, we speak on the exterior horizon only).
However, the tetrad given above is inadmissible already on the
surface of the static limit (which suggests the interpretation of
the latter). Below this limit all four tetrad (co)vectors are
spacelike, and only under the horizon the tetrad covector
$\theta^{(1)}$ can play the r\^ole of timelike congruence which is
however non-rotating (instead we observe pseudo-rotation of
$\theta^{(0)}$, now being spacelike). Of course, in the region
between the surfaces of static limit and horizon (excluding the
horizon itself) one can still use easily normalizable timelike
combinations of $\theta^{(0)}$ and $\theta^{(3)}$ as the new 0th
tetrad covector, though it always serves only in a final radial
region (the so-called `local stationarity' of the Kerr--Newman
space-time in the ergosphere). Thus it seems that there is an
abrupt change from rotation to pseudo-rotation when the horizon is
being crossed, but this is not exactly the case: since the
Boyer--Lindquist coordinates at the horizon are inadmissible, the
exterior and interior (with respect to horizon) space-time regions
are absolutely disjoint. The only way to deal with this problem is
to introduce a system of synchronous coordinates which, however,
does not rotate {\em per se}.

{\em The pseudo-rotating frame.} Another combination of terms in
the Kerr metric in Boyer--Lindquist coordinates yields the
pseudo-rotating (but not rotating in the sense of the timelike
congruence of $\theta^{(0)}$) orthonormal basis (the notations are
now changed in all cases, essentially with the exception of
$\Delta$ and $\rho$) \be \label{KNpsr} \theta^{(0)}=e^\alpha dt, ~
\theta^{(1)}=e^\beta dr, ~ \theta^{(2)}=e^\gamma d\vartheta, ~
\theta^{(3)}=e^\delta \sin\vartheta(d\phi+aFdt), \ee
$$
e^{2\alpha}=\frac{\rho^2\Delta}{(r^2+ a^2)^2-\Delta
a^2\sin^2\vartheta}, ~ e^{2\beta}= \frac{\rho^2}{\Delta}, ~
e^{2\gamma}=\rho^2,
$$
$$
e^{2\delta}=\frac{(r^2+a^2)^2-\Delta a^2\sin^2\vartheta}{\rho^2}.
$$
Here
$$
F=a\frac{r^2+a^2-\Delta}{(r^2+a^2)^2 -\Delta a^2\sin^2\vartheta}
$$
is a new function, though similar to $f(r,\vartheta)$ of the
preceding basis choice. At $r=Q^2/2M$ the pseudo-rotation vanishes
(it changes direction when crossing this sphere); the same occurs
with the rotation in the preceding basis. The equation
$(\rho^2+a^2)^2-\Delta a^2\sin^2\vartheta=0$ has no real solutions
for realistic values of the charge $Q$. This means that the only
singularity of the pseudo-rotating tetrad occurs at the horizon
($\Delta=0$), but this is a singularity of the basis (and of the
system of coordinates) only.

The Kerr--Newman electromagnetic stress-energy tensor is related
to the rotating tetrad $\theta^{(\alpha)}$ as to its proper basis;
the pseudo-rotating tetrad is not built of eigenvectors of the
electromagnetic field.

\vspace*{1.cm}

\noindent{\large\bf 5. A PSEUDO-ROTATING SPACE-TIME}

\vspace*{0.5cm}

\noindent An example of a pseudo-rotating space-time was found in
the electrovacuum case in [1] (see also [4], p. 222; but this
result was not included in the new edition [13]). In this `static'
space-time the orthonormal covector basis is \be \label{Ch} \left.
\begin{array}{cc} \theta^{(0)}= b\rho^{-2/9}e^{(1/2)
a^2\rho^{2/3}}dt, & \theta^{(1)}=b\rho^{-2/9}e^{(1/2)a^2
\rho^{2/3}}d\rho,\\ \theta^{(2)}=\rho^{2/3}d\phi, & \theta^{(3)}
=\rho^{1/3}\left(dz+a\rho^{2/3}d\phi\right), \end{array} \right\}
\ee being accompanied by the sourceless (outside the symmetry
axis) electromagnetic field with the potential 1-form \be
\label{Chem} A=\frac{4a}{3} \sqrt{\frac{\pi}{\varkappa}}\phi dt=-
\frac{4a}{3} \sqrt{\frac{\pi}{\varkappa}}td\phi+\mbox{ exact
form}\ee and the stress-energy tensor \be \label{ChT}
T=\frac{2a^2}{9\varkappa b^2}
\rho^{-8/9}e^{-a^2\rho^{2/3}}\left(\theta^{(0)}\otimes\theta^{(0)}
+\theta^{(1)}\otimes\theta^{(1)}-\theta^{(2)}\otimes\theta^{(2)}+
\theta^{(3)}\otimes\theta^{(3)}\right). \ee The tetrad basis
(\ref{Ch}) obviously is a proper one for the electromagnetic field
stress-energy tensor (\ref{ChT}), thus this source well matches
the property of $\theta^{(3)}$-pseudo-rotation and {\em vice
versa}. The electromagnetic field is either of purely electric or
purely magnetic type (one case is merely the dual conjugate of the
other), the latter permitting a more natural physical
interpretation. Then the magnetic covector is the only non-trivial
one in the above basis and directed along $\varphi$: \be B=\frac{4
\sqrt{\pi}a\exp(-(a^2/2)\rho^{2/3})}{3b\sqrt{\varkappa}
\rho^{10/9}}\theta^{(2)}. \ee

\vspace*{1.cm}

\noindent{\large\bf 7. CONCLUDING  REMARKS}

\vspace*{0.5cm}

In this paper we have seen that the rotation phenomenon has very
different sides: {\bf(a)} in general, it separates in two
alternative cases, proper rotation and pseudo-rotation, or their
combination, rotation being related to timelike vectors, and
pseudo-rotation, to spacelike ones; {\bf(b)} this phenomenon can
be related to rotating congruence(s) and rotating tetrad(s),
leaving its mark primarily upon metric; {\bf(c)} from the
viewpoint of the material contents of the space-time, it
corresponds to rotation of eigenvector(s) of the stress-energy
tensor; {\bf(d)} if isometries are taken into account, rotating
Killing vector field(s) should be considered, and this is the only
method to locally deal with this phenomenon in a vacuum; {\bf(e)}
in a perfect fluid, the rotation is considered as that of the
fluid's four-velocity vector field, and in the $r$-form field
theoretic description of fluids it is equivalent to a presence of
inhomogeneity term in the dynamic field equation (which however
cannot be interpreted as a source term, in contrast to the
traditional treatment of such terms in the gravitational and
electromagnetic equations). It is interesting that more than one
timelike or spacelike Killing vectors with rotation or
pseudo-rotation, respectively, may exist simultaneously (for
example, there are even four rotating independent timelike Killing
vectors in the G\"odel space-time, see [11]) .

In the concrete examples of rotation and pseudo-rotation, we used
here the electromagnetic field as a material contents of
space-time, since in general relativity this field proved to be
much richer of sufficiently simple and informative exact
self-consistent solutions than any other type of distributed
sources. Considering these examples, we not only illustrated the
different sides of the rotation phenomenon, in particular, showing
that pseudo-rotation frequently is an indispensable counterpart of
rotation, but we also have drawn some new conclusions about the
geometrical and physical properties of a specific choice of exact
solutions: {\bf 1.} The special case of conformally flat null
Einstein--Maxwell field (Section 3) admits seven independent
Killing vectors, exclusively in terms of four of which its metric
can be expressed. {\bf 2.} This is in fact a multitude of
self-consistent exact solutions with radically different null
electromagnetic fields, but with one and the same space-time
geometry, while several Killing vectors of this space-time serve
as four-potentials for these electromagnetic fields (not merely
test ones, as this could be normally expected). The symmetry group
of the electromagnetic field is more restricted that that of the
resulting space-time (semi-cylindrical symmetry \`a la Wils). {\bf
3.} In the Kerr--Newman space-time (Section 4) it was shown that
the (usual) rotating tetrad becomes pseudo-rotating inside the
event horizon, still being built of eigenvectors of the
electromagnetic stress-energy tensor, and the properties of the
pseudo-rotating (but not rotating) tetrad outside the horizon were
studied.

\vspace*{1.cm}

\noindent{\large\bf ACKNOWLEDGMENTS}

\vspace*{0.5cm}

\noindent This paper contains a part of H. V.-R.'s Ph. D. Thesis,
and the CONACyT-M\'exico scholarship grant No. 91290 is gratefully
acknowledged. We thank Marcello Ortaggio for valuable critical
remarks.

\vspace*{1.cm}

\noindent{\large\bf REFERENCES}\footnotesize{

\vspace*{0.5cm}

\noindent 1. Chitre, D. M., G\"uven, R., and Nutku, Y. (1975).
{\it J. Math. Phys.} {\bf 16}, 475.

\noindent 2. Fayos, F., and Sopuerta, C. F. (1999). {\it Class.
Quantum Grav.} {\bf 16}, 2965.

\noindent 3.  Horsk\'y, J., and Mitskievich, N. V.(1989). {\it
Czech. J. Phys.} {\bf B39}, 957.

\noindent 4. Kramer, D., Stephani, H., MacCallum, M., and Herlt,
E. (1980). {\it Exact Solutions of \phantom{aaa}Einstein's Field
Equations}, Cambridge, UK: Cambridge University Press.

\noindent 5. Lichnerowicz, A. (1955). {\it Th\'eories Relativistes
de la Gravitation et de l'\'Electromagn\'et\-\phantom{aaa}isme},
Paris: Masson et Cie.

\noindent 6. Lightman, A. P., Press, W. H., Price, R. H., and
Teukolsky, S. H. (1975). {\it Problem \phantom{aaa}Book in
Relativity and Gravitation}, Princeton, NJ: Princeton University
Press.

\noindent 7. Misner, C. W., Thorne, K. S., and Wheeler, J. A.
(1973). {\it Gravitation}, San Francisco: \phantom{aaa}W. H.
Freeman.

\noindent 8. Mitskievich, N. V. (1996). {\it Relativistic Physics
in Arbitrary Reference Frames}, gr-\phantom{aaa}qc/9606051.

\noindent 9. Mitskievich, N. V. (1999). {\it Int. J. Theor. Phys.}
{\bf 38}, 997.

\noindent 10. Mitskievich, N. V. (1999). {\it Gen. Rel Grav.} {\bf
31}, 713.

\noindent 11. Mitskievich, N. V. (2001). Lorentz force free
charged fluids in general relativity: \phantom{aaa}The physical
interpretation. In: {\it Exact solutions and Scalar Fields in
Gravity: Recent \phantom{aaa}Developments}, New York: Kluwer
Academic/Plenum Publishers, p. 311.

\noindent 12. Mitskievich, N. V. (2003). {\it Rev. Mex. de
F{\'\i}sica} {\bf 49 Supl. 2}, 39; (2002). {\it Spacetimes,
\phantom{aaa}electromagnetism and fluids (a revision of
traditional concepts)}, gr-qc/0202032.

\noindent 13. Stephani, H., Kramer, D.,  MacCallum, M.,
Hoenselaers, C., and Herlt, E. (2003). \phantom{aaa}{\it Exact
Solutions of Einstein's Field Equations, Second Edition},
Cambridge, UK: Cam\-\phantom{aaa}bridge University Press.

\noindent 14. Synge, J.L. (1965). {\it Relativity: The Special
Theory}, Amsterdam: North-Holland. }

\noindent 15. Mitskievich, N.V., and Vargas Rodr\'iguez, H. (2005)
{\it General Relativity and Gravitation}, {\bf 37}, No. 4, 781.

\end{document}